\documentclass{appolb}
\usepackage{epsfig}

\begin{document}
\pagestyle{plain}
\newcount\eLiNe\eLiNe=\inputlineno\advance\eLiNe by -1
\title{DEEPLY VIRTUAL COMPTON SCATTERING\\
AND PROMPT PHOTON PRODUCTION\\
AT ZEUS AND H1 EXPERIMENTS%
}
\author{Xavier JANSSEN
\address{DESY, Notkestr. 85, D-22607 Hamburg, Germany}}
\maketitle

\begin{abstract}
Recent results on prompt photon production and Deeply Virtual Compton
Scattering (DVCS) from ZEUS and H1 experiments are presented.
\end{abstract}

\section{Deeply Virtual Compton Scattering}

\begin{figure}[b]
 \begin{center}
  \vspace*{-0.3cm}
  \epsfig{figure=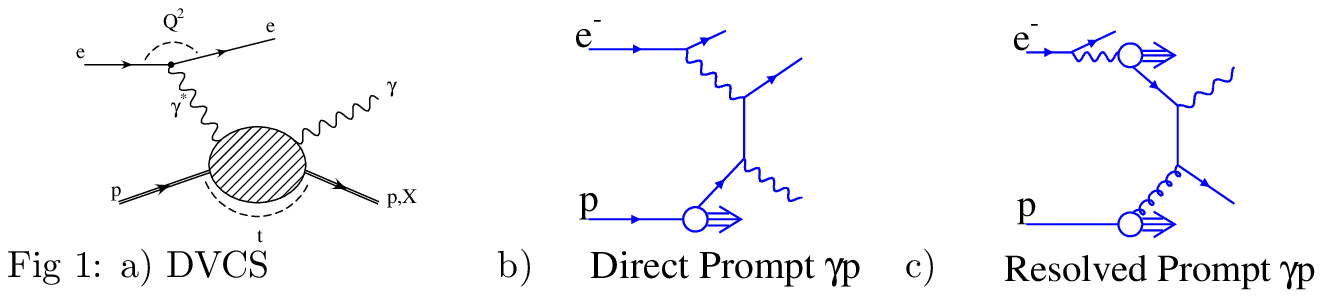,width=0.90\textwidth}
  \vspace*{-0.8cm}
  \addtocounter{figure}{1}
   \label{fig:diag}
 \end{center}
\end{figure}

The interest of Deeply Virtual Compton Scattering (DVCS), i.e. exclusive 
photon production off the proton $\gamma^{*} p \longrightarrow \gamma p$ 
(see Fig. 1a), is in the insight it gives to the applicability of
perturbative Quantum Chromo Dynamics (QCD) in the field of diffractive
interactions and to the nucleon partonic structure. The hard interaction 
between the photon and the proton proceeds via the exchange of at least 
two quarks at leading order or two gluons at next-to-leading order
in a colour neutral state. The transition from a virtual to a real photon
forces the fractional momenta of the two partons involved to be different
("skewed"). To take into account this skewing effect, one has to consider
generalised parton distributions (GPDs) which are an extension of the
standard parton densities (PDFs) including additional information on the correlations
between partons and their transverse motion. 
The measurements presented here are compared with a NLO QCD prediction
in which the DVCS cross section has been calculated using 
GPD parametrisations \cite{Freund:2002qf} based on the
ordinary PDFs in the DGLAP formalism and where
the skewedness is dynamically generated. 
The $t$ dependence of the GPDs, 
where $t$ is the square of the four-momentum exchanged at the proton vertex, 
is taken to be $e^{-b|t|}$.

This paper presents measurements of DVCS cross sections performed by 
the H1 \cite{h1dvcs97} and the ZEUS \cite{zeusdvcs} experiments. 
The DVCS cross section has been measured differentially
in $t$ by the H1 collaboration for two different values of the
photon virtuality $Q^2$ as shown in Fig. 2. The observed fast decrease
with $|t|$ can be described by the form $e^{-b|t|}$. Combining  
data from both  $Q^2$ range, the $|t|$ slope is measured to be $b = 6.02 \pm 0.35 \rm{(stat.)}
\pm 0.39 \rm{(syst.)} \ \rm{GeV}^{-2}$ at $Q^2 = 8 \ \rm{GeV}^2$
and $W = 82 \ \rm{GeV}$. 
The measurement constrains models
as their normalisation depends directly on the $t$ slope parameter.

\begin{figure}
 \begin{center}
  \epsfig{figure=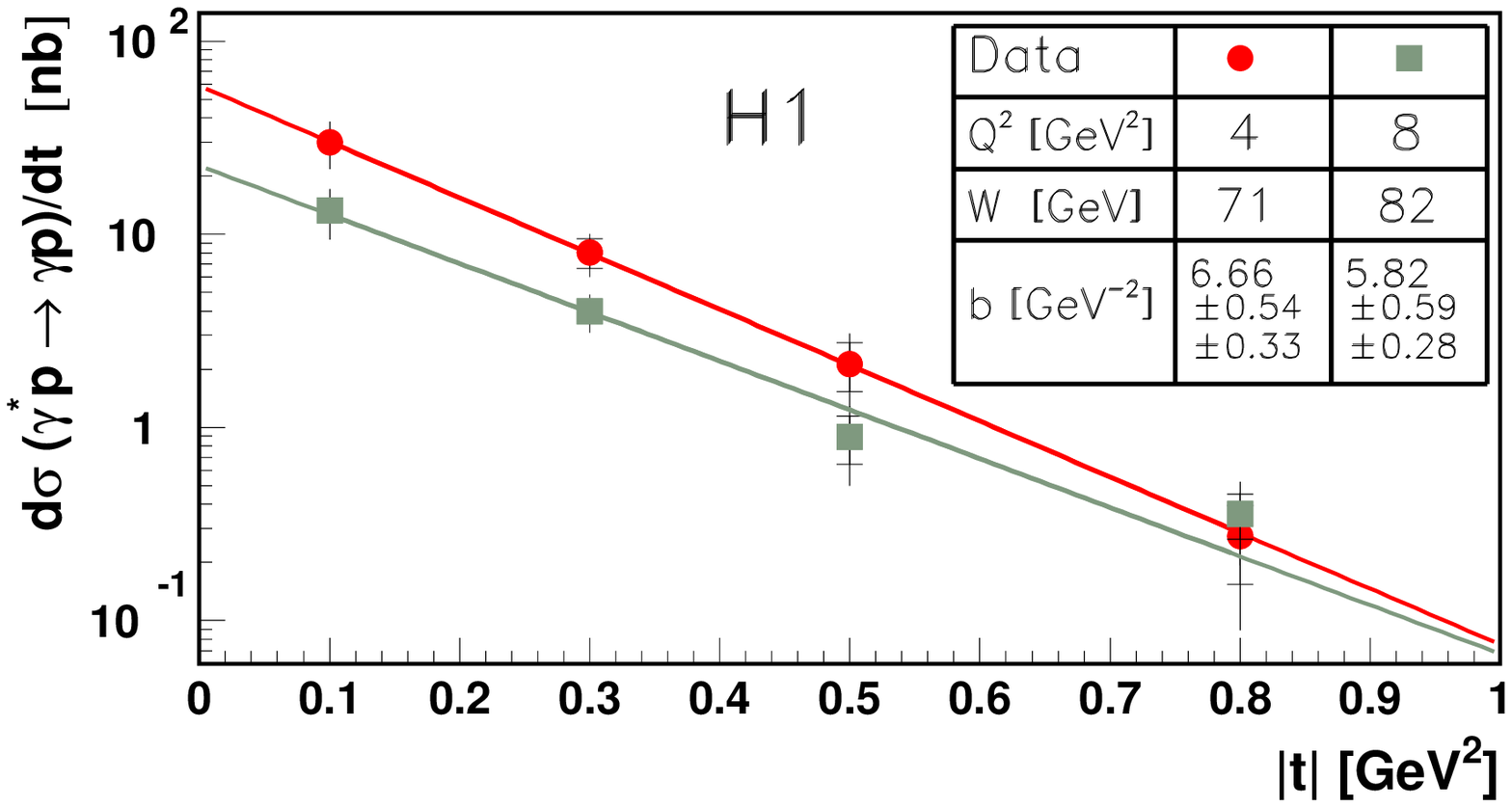,width=0.55\textwidth}
  \begin{picture}(5.0,0.)
    \put(-80,56) {\epsfig{file=janssen_fig2b.eps,width=2.55cm,height=0.63cm}}
  \end{picture}
 \end{center}
 \vspace*{-0.4cm}
 \caption{\sl The cross section $\gamma^* p \rightarrow \gamma p$
  differential in $t$, for $Q^2=4$~GeV$^2$ and $Q^2=8$~GeV$^2$.
  The inner error bars represent the statistical and the
  full error bars the quadratic sum of the statistical and systematic
  uncertainties.}
 \label{fig:dsdt}
 \vspace*{-0.4cm}
\end{figure}

\begin{figure}[htb]
 \begin{center}
  \epsfig{figure=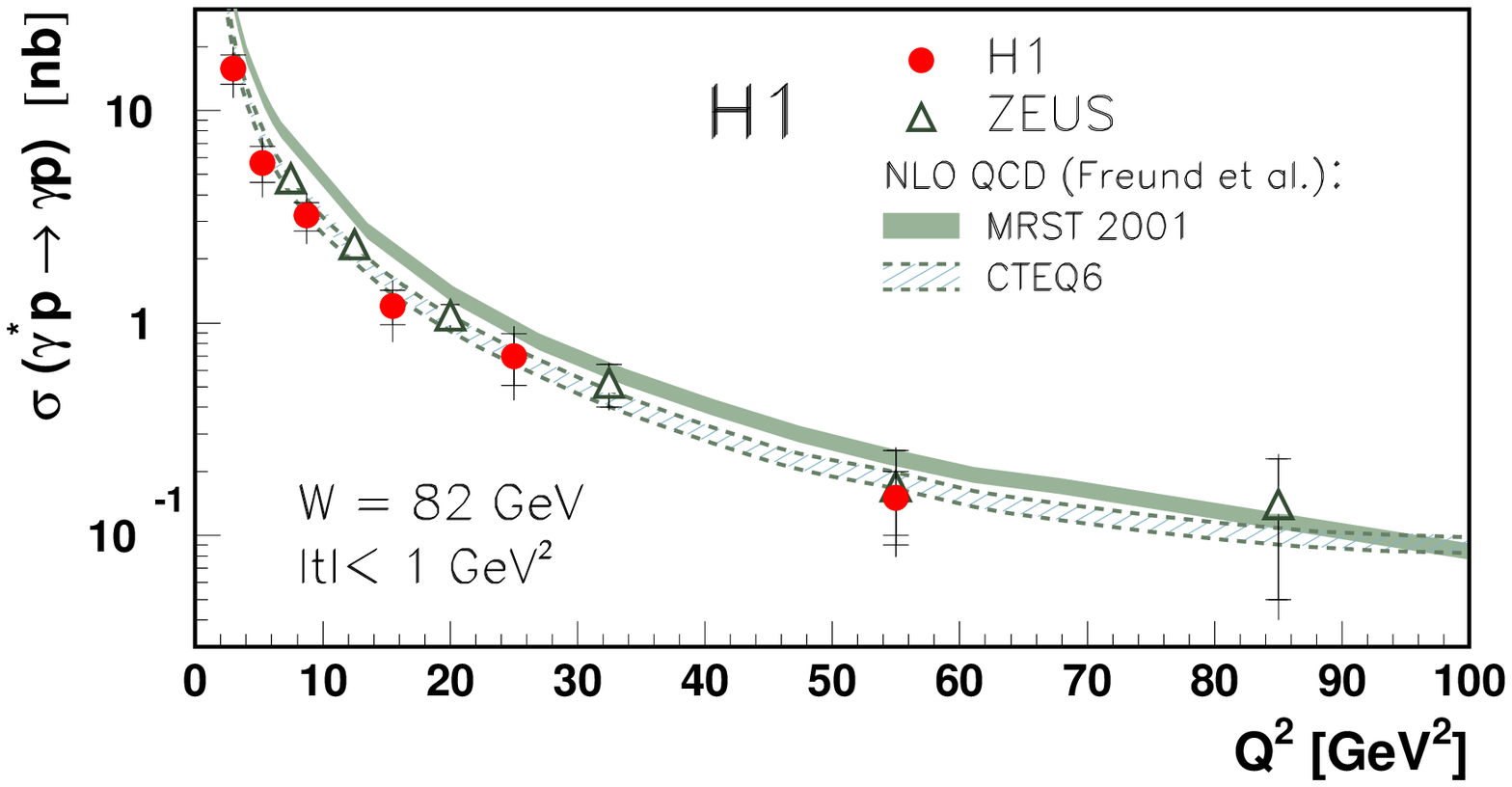,width=0.49\textwidth}
  \epsfig{figure=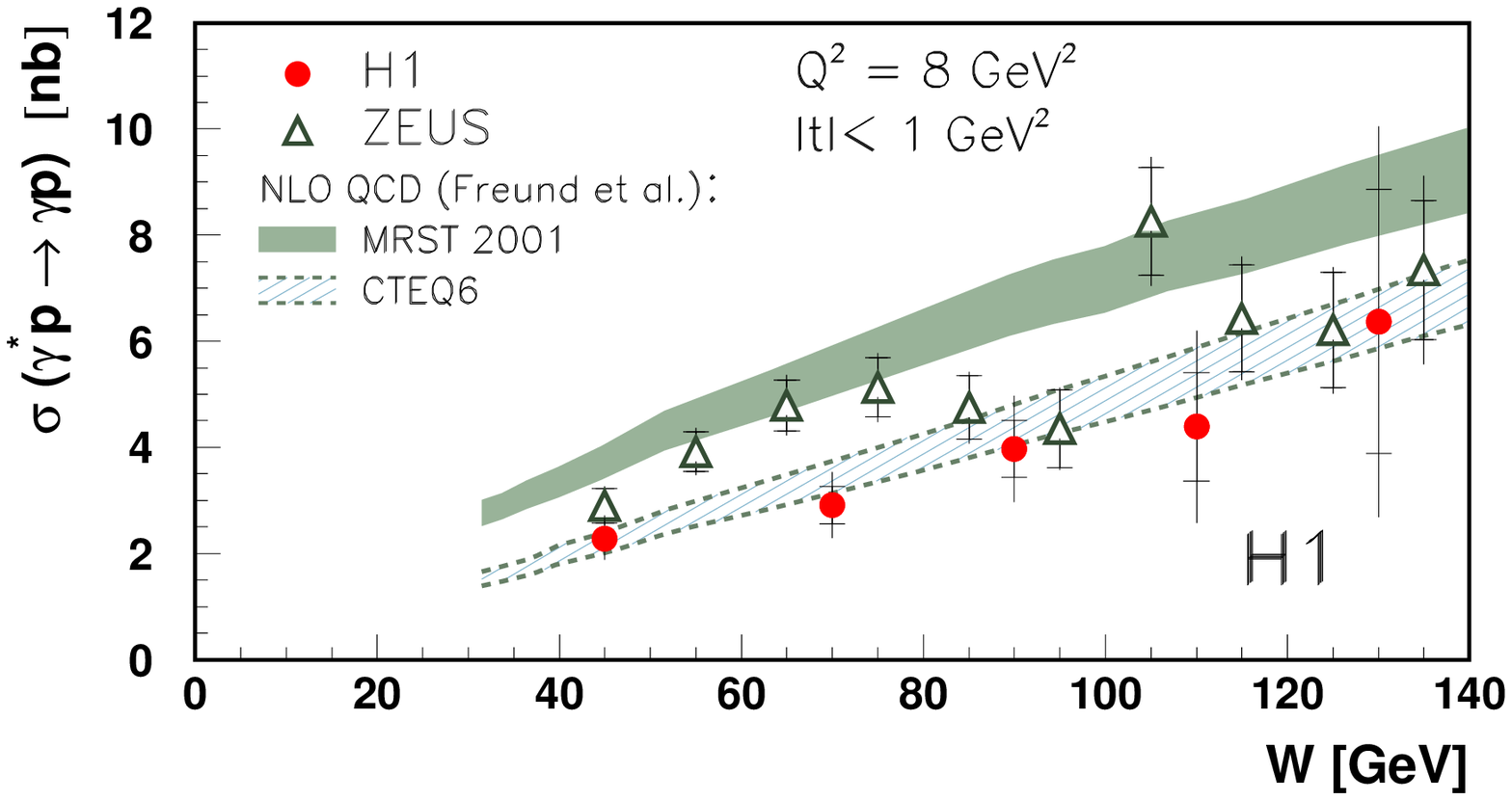,width=0.49\textwidth}
  \vspace*{-0.4cm}
 \end{center}
 \caption{\sl The $\gamma^* p \rightarrow  \gamma p$ cross section
  as a function of $Q^2$ for $W=82$ GeV (left) and 
  as a function of $W$ for $ Q^2=8$ GeV$^2$ (right).
  The H1 measurement is shown together with the results of
  ZEUS and NLO QCD predictions based on MRST 2001
  and CTEQ6 PDFs.
  The band associated with each prediction corresponds to the
  uncertainty on the measured $t$-slope.
  }
 \label{fig:sqcd}
\end{figure}

The $\gamma^{*} p \longrightarrow \gamma p$ cross section is presented
in Fig. 3 as a function of $Q^2$ and the photon-proton centre-of-mass
energy $W$. The $W$ dependence can be parametrised as $\sigma \propto 
W^{\delta}$ yielding $\delta = 0.77 \pm 0.23 \pm 0.19$ at $Q^2=8$ $\rm{GeV}^2$, 
which is similar to J/$\Psi$ production \cite{jpsi} indicating 
a hard scattering process. Fig. 3 also compares the measurements
with the NLO QCD predictions \cite{Freund:2002qf} where the 
PDFs of MRST2001 and CTEQ6 are used as input for the GPDs. 
The theoretical estimates
agree well with the data both in shape and absolute normalisation.
Reducing further the experimental errors will set constraints
on different GPDs models.

\section{Prompt Photon Production}

Prompt photons in the final state of high energy collisions 
(Fig. 1b and 1c) provide a detailed study of perturbative QCD.
The term ``prompt" refers to photons which are radiated directly 
from partons of the hard interaction, instead of stemming from
the decay of hadrons or from QED radiation from the electrons.
In contrast to jets, prompt photons are not affected by
  hadronisation resulting in the reduction of this theoretical
  uncertainty.
The main experimental difficulty
is the separation of the prompt photons from hadronic background,
in particular from signals due to $\pi^0$ mesons. Separation is 
performed on basis of properties of the energy deposit structure in
the calorimeters of the H1 and ZEUS experiments.

\begin{figure}[h] 
  \vskip 1.0cm
 \begin{picture}(100,100)(0,0)
 \put(-10,0){\epsfig{figure=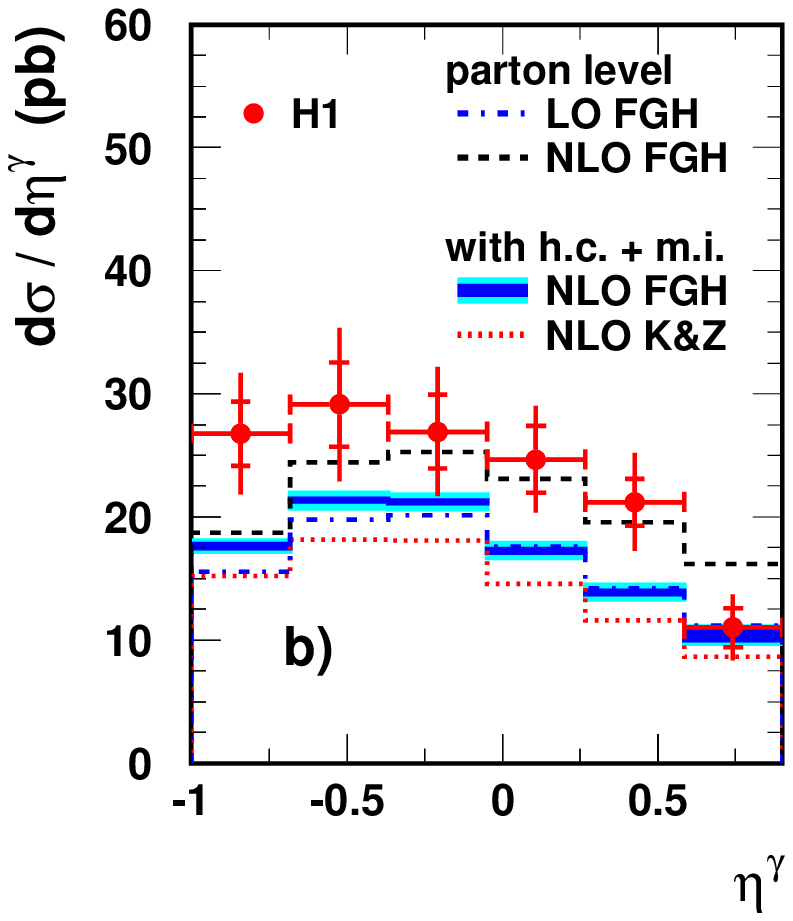,width=0.32\textwidth}}
 \put(110,0){\epsfig{figure=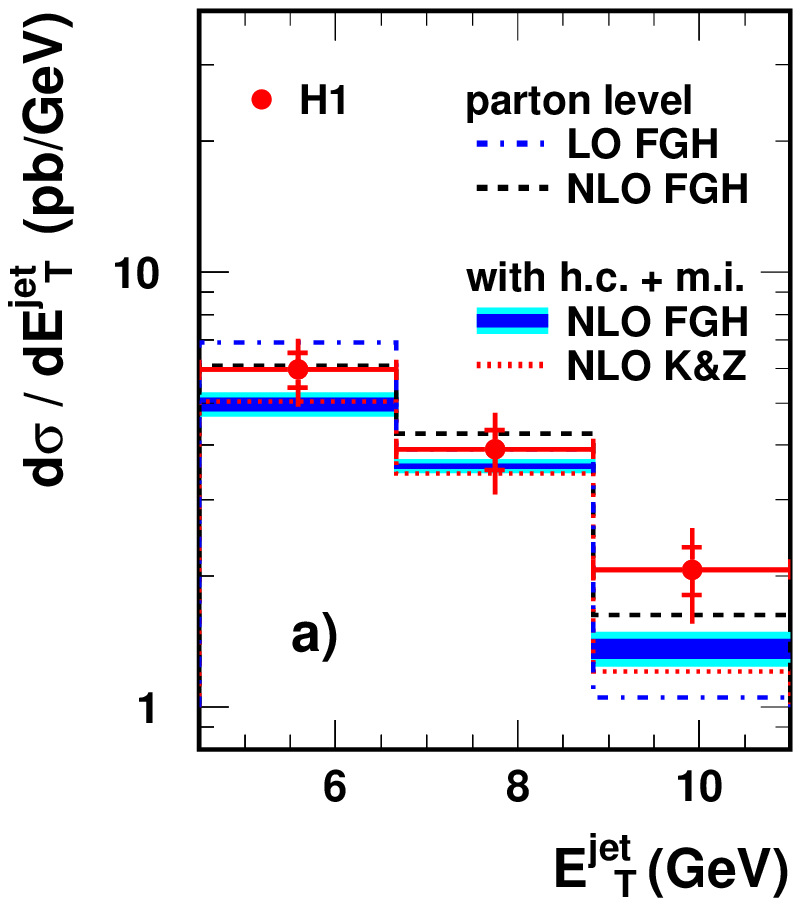,width=0.32\textwidth}}
 \put(240,13){\epsfig{figure=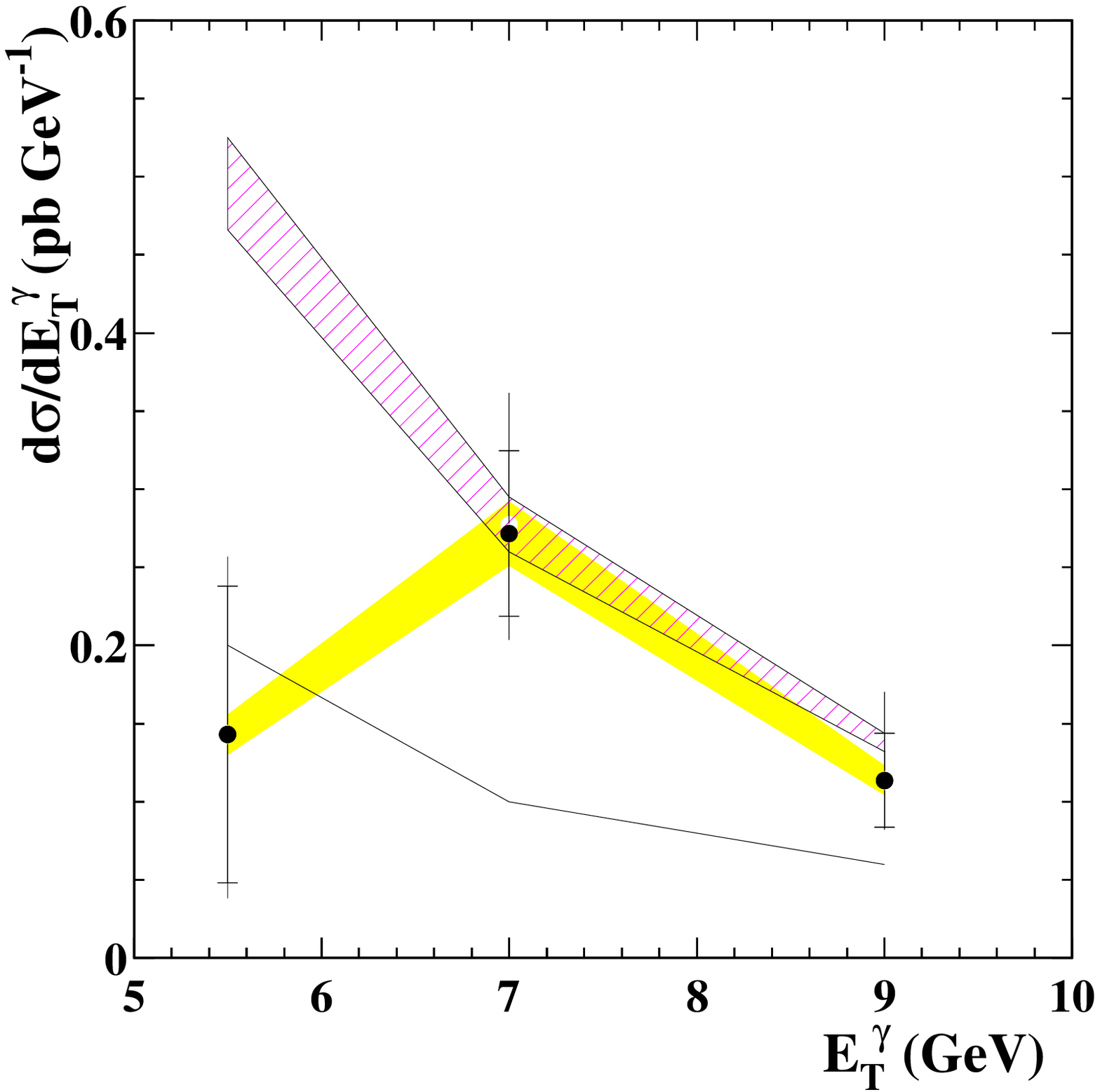,width=0.32\textwidth}}
 \put(330,110){c)}
 \put(285,110){\small ZEUS}
 \end{picture}
 \vspace*{-0.4cm}
 \caption{Prompt photon differential cross sections.
   a) in inclusive photoproduction,
   b) in photoproduction with a jet,
   c) in DIS with a jet - the hatched band represents the NLO QCD prediction and the solid line the contribution from QED radiation.
  }
 \label{fig:prompt}
\end{figure}

Fig. 4a shows the differential inclusive prompt photon cross section
as a function of the photon pseudorapidity $\eta^{\gamma}$ in the
photoproduction regime ($Q^2 < 1 \ \rm{GeV}^2$ and $142 < W < 266 \ \rm{GeV}$)
as measured by H1 \cite{Aktas:2004uv}. Similar results have been obtained
by the ZEUS experiments \cite{Breitweg:1999su}. The data are compared to
NLO perturbative QCD calculations by Fontannaz, Guillet and Heinrich 
(FGH) \cite{Fontannaz:2001ek} and by Krawczyk and Zembrzuski (K\&Z)
\cite{Krawczyk:2001tz} which both provide a good description in shape
but are 20-30\% lower. The cross section for
prompt photons when an additional jet with $E_T^{jet} > 4.5 \ \rm{GeV}$
and $-1 < \eta_{jet} < 2.3$ is required is shown on Fig. 4b as a 
function of the jet transverse energy $E_T^{jet}$. Both NLO perturbative QCD 
calculations \cite{Fontannaz:2001nq,Zembrzuski:2003nu} provide a good
description of the data in shape and normalisation.

The inclusive prompt photon cross section in DIS ($Q^2 > 35 \ \rm{GeV}^2$)
with a jet ($E_T^{jet} > 6 \ \rm{GeV}$ and $-1.5 < \eta_{jet} < 1.8$) as
measured by the ZEUS collaboration \cite{Chekanov:2004wr} is shown in Fig. 4c
versus the prompt photon transverse energy $E_T^{\gamma}$. The data are 
compared to a NLO QCD calculation \cite{Gehrmann-DeRidder:2000ce}
which includes QED radiative corrections 
 on the electron lines and a good 
description is obtained except at low $E_T^{\gamma}$ and in the more
forward (outgoing proton beam) direction. Without QED radiative corrections,
the NLO QCD calculation would undershoot the
data.

The HERWIG and PYTHIA Monte Carlo predictions undershoot the data both in
photoproduction and DIS cases, see \cite{Aktas:2004uv,Chekanov:2004wr}
for more details.

\begin{flushright}
{\small
The author is also supported by the {\it Fonds National \\
de la Recherche Scientifique} of Belgium.}
\end{flushright}

\end{document}